\documentclass[12pt]{article}
\usepackage{amssymb}
\addtocounter{page}{0} \tolerance=5000
\newcommand{\beq}{\begin{equation}}
\newcommand{\eeq}{\end{equation}}
\newcommand{\beqn}{\begin{eqnarray}}
\newcommand{\eeqn}{\end{eqnarray}}

\newcommand{\p}{\mbox{${\bf p}$}}

\newcommand{\s}{\mbox{${\bf s}$}}

\newcommand{\bk}{\mbox{${\bf k}$}}
\newcommand{\bR}{\mbox{${\bf R}$}}
\newcommand{\br}{\mbox{${\bf r}$}}

\newcommand{\si}{\mbox{{\boldmath$\sigma$}}}

\newcommand{\al}{\mbox{${\alpha}$}}
\newcommand{\De}{\mbox{${\Delta}$}}

\begin{document}
\begin{titlepage}
\begin{center}
{\bf Corrections to Deuterium Hyperfine Structure\\ Due to
Deuteron Excitations}
\end{center}

\begin{center}
I.B. Khriplovich\footnote{e-mail address: khriplovich@inp.nsk.su},
A.I. Milstein\footnote{e-mail address: milstein@inp.nsk.su}\\
Budker Institute of Nuclear Physics\\ and Novosibirsk
University,\\ 630090 Novosibirsk, Russia
\end{center}

\bigskip

\begin{abstract}
We consider the corrections to deuterium hyperfine structure
originating from the two-photon exchange between electron and
deuteron, with the deuteron excitations in the intermediate
states. In particular, the motion of the two intermediate nucleons
as a whole is taken into account. The problem is solved in the
zero-range approximation. The result is in good agreement with the
experimental value of the deuterium hyperfine splitting.
\end{abstract}

\vspace{7cm}

\end{titlepage}

\section{Introduction}
 The hyperfine (hf) splitting in deuterium ground state
has been measured with high accuracy. The most precise
experimental result for it was obtained with an atomic deuterium
maser and constitutes \cite{wr}
\begin{equation}
\nu_{\rm exp}\,=\,327\,384.352\,522\,2(17)\, \mbox{kHz}.
\end{equation}
Meanwhile the theoretical calculation including higher order pure QED
corrections gives
\begin{equation}
\nu_{\rm QED}\,=\,327\,339.27(7)\,
\mbox{kHz}.\;\;\;\;\;\;\;\;\;\;\;\;\;\;\;
\end{equation}
The last number was obtained by using the theoretical result for
the hydrogen hf splitting from Ref. \cite{by}
$$1\,420\,451.95(14)\,\mbox{kHz}$$ which does not include proton
structure and recoil radiative correction, and combining it with
the theoretical ratio of the hf constants in hydrogen and
deuterium from Ref. \cite{ra} $$4.339\,387\,6(8)$$ based on the
ratio of the nuclear magnetic moments and including the reduced
mass effect in~$|\psi(0)|^2$.

It was recognized long ago that the discrepancy
\begin{equation}\label{dif}
\nu_{\rm exp}\,-\,\nu_{\rm QED}\,=\,45\, \mbox{kHz}
\end{equation}
is due to the effects caused by the finite size of deuteron. Such
effects are obviously much larger in deuterium than in hydrogen.
Corresponding contributions to the deuterium hf splitting were
discussed long ago with some intuitive arguments \cite{bo}, and
then in more detail in Refs. \cite{low,los,gf}.

We believe that in the past the most systematic treatment of such
effects, which are due to the electron-deuteron interaction of
second order in $\al$, was performed in Ref. \cite{mkp}. The
effective hamiltonian of the hf interaction of second order in
$\alpha=e^2/4\pi$ was derived therein from the elastic forward
scattering amplitude of virtual photons off the deuteron.

In particular, the low-energy theorem for forward Compton
scattering \cite{lo,gg,fe,mi} was generalized in \cite{mkp} to the
case of virtual photons and a target with arbitrary spin. The
corresponding contribution of the momentum transfers $k$, bounded
from above by the inverse deuteron size $\varkappa\,=\,45.7$ MeV,
to the relative correction to the deuterium hyperfine structure is
\begin{equation}\label{de}
\Delta_{el}^d=\,\frac{3\alpha}{8\pi} \left(\mu_d
-2-\,\frac{3}{\mu_d}\right)\frac{m_e}{m_p}\,\ln{\frac{\varkappa}{m_e}}.
\end{equation}
Here $m_e$ and $m_p$ are the electron and proton masses,
respectively, $\mu_d=0.857$ is the deuteron magnetic moment. The
relative corrections $\De$'s are defined here and below as the
ratios of the corresponding contributions to the $ed$ scattering
amplitude to the spin-dependent Born term in this amplitude,
\begin{equation}\label{b}
T_{0}=\,-\,\frac{2\pi\alpha}{3m_e m_p}\, \mu_d\,(\si \cdot\s),
\end{equation}
where $\s$ is the deuteron spin.

At larger momentum transfers, $k>\varkappa$, the amplitude of the
Compton scattering on a deuteron is just the coherent sum of those
amplitudes on free proton and neutron. This correction equals
\begin{equation}\label{di}
\Delta_{in}^{pn}=\,\frac{3\alpha}{4\pi}\,\frac{1}{\mu_d}\,
(\,\mu_p^2-\,2\,\mu_p\,-\,3\,+\,\mu_n^2\,)\,\frac{m_e}{m_p}\,
\ln{\frac{m_{\rho}}{\varkappa}}.
\end{equation}
Here $\mu_p\,=\,2.79$ and $\mu_n\,=\,-\,1.91$ are the proton and
neutron magnetic moments; $m_{\rho}~=~770$~MeV is the usual
hadronic scale.

Strong numerical cancellation between $\Delta_{el}$ and
$\Delta_{in}$ is worth mentioning.

The next correction to the deuterium hyperfine structure (hfs),
obtained in Ref. \cite{mkp}, is induced by the deuteron virtual
excitations due to spin currents only. It is
\begin{equation}\label{di1}
\Delta_{in}^{(1)}=\,\frac{3\alpha}{8\pi}\,
\frac{(\,\mu_p\,-\,\mu_n\,)^2}{\mu_d}\,\frac{m_e}{m_p}\,\ln{\frac{m_p}{\varkappa}}\,.
\end{equation}

There is also a correction due to a finite distribution of the
deuteron charge and magnetic moment\footnote{In the case of
hydrogen this problem was considered many years ago by Zemach
\cite{ze}.}. In the zero-range approximation used in Ref.
\cite{mkp} this correction is
\begin{equation}\label{zem}
\Delta_{f}=\,-\,\alpha\,\frac{m_e}{3\,\varkappa}\,(1+2\ln2).
\end{equation}

\section{Leading inelastic nuclear correction\\ to deuterium hyperfine structure}

Leading inelastic nuclear correction is on the relative order
$\alpha\,m_e/\varkappa$ (as well as the Zemach correction
(\ref{zem})). The corresponding effect calculated in Ref.
\cite{mkp} is additionally enhanced by a large factor
$\mu_p\,-\mu_n\,=4.7$. In the present paper we consider two more
effects on the same order $\alpha\,m_e/\varkappa$. Though both of
them are proportional to $\mu_p\,+\mu_n\,=0.88$ (and thus are
essentially smaller numerically than that considered in
\cite{mkp}), we believe that their investigation is worth
attention.

We use the gauge $A_0 =\,0$ where the photon propagator is
\begin{equation}\label{po}
D_{im}(\omega,
\vec{k})\,=\,\frac{\,d_{im}}{\omega^2-\,\vec{k}^2}\,,\;\;
d_{im}=\,\delta_{im}-\,\frac{k_i k_m}{\omega^2}\,;\;\;\;
D_{00}=\,D_{0m}=\,0.
\end{equation}
The electron-deuteron nuclear-spin-dependent scattering amplitude
generated by the two-photon exchange is
\begin{equation}\label{f}
T =\,4\pi\alpha i\,\int\frac{d^4
k}{(2\pi)^4}\,\frac{d_{im}d_{jn}}{k^4}\,
\frac{\gamma_i(\hat{l}-\hat{k}+m_e)\gamma_j}{k^2-\,2lk}\, M_{mn}.
\end{equation}
Here $l_{\mu}=\,(m_e,0,0,0)$ is the electron momentum. The
structure $\gamma_i(\hat{l}-\hat{k}+m_e)\gamma_j$ reduces to
$-\,i\,\omega\,\epsilon_{ijl}\sigma_l$ where $\vec{\sigma}$ is the
electron spin. We calculate the nuclear matrix elements entering
the deuteron Compton amplitude $M_{mn}$ in the zero-range
approximation (zra) which allows us to obtain all the results in a
closed analytical form.

The inelastic $1/\varkappa$ contribution to hfs is induced by the
combined action of the convection and spin currents. Since the
convection current is spin-independent, all the intermediate
states are triplet ones, as well as the ground state. Therefore,
here the spin current operator
\[
\frac{e}{2m_p}\,i\bk \times [\mu_p \si_p \exp(i\bk
 \br_p)+ \mu_n \si_n \exp(i\bk \br_n)]
\]
simplifies to
\begin{equation}\label{si}
\frac{e}{2m_p}\,i[\bk \times \s]\; [\mu_p  \exp(i\bk
 \br_p)+
\mu_n \exp(i\bk \br_n)].
\end{equation}
In the initial state $|\,0 \rangle$ the deuteron is at rest. But
in the excited state the system of nucleons as a whole moves with
the momentum $\bk$, so that its wave function is $|\,n \rangle
\exp(i\bk \bR)$, where $|\,n \rangle$ refers to the deuteron
internal degrees of freedom and is a function of $\br = \br_p -
\br_n$; $\bR = (\br_p + \br_n)/2$ is the deuteron centre of mass
coordinate. Thus, a typical matrix element of the spin current can
be written as
\[
\frac{e}{2m_p}\,i\bk \times \langle n|\exp(-i\bk \bR) \s [\mu_p
\exp(i\bk \br_p)+ \mu_n \exp(i\bk \br_n)]|0\rangle\,
\]
\beq\label{sc}
=\frac{e}{2m_p}\,i\bk \times \langle n|\s [\mu_p \exp(i\bk \br/2)+
\mu_n \exp(-i\bk \br/2)]|0\rangle\,.
\eeq
As to a typical matrix element of the convection current, it
transforms now as follows:
\[
\frac{e}{2m_p}\,\langle n|\exp(-i\bk \bR) \hat{\p}_p \exp(i\bk
\br_p)+ \exp(i\bk \br_p)\hat{\p}_p|0\rangle\,
\]
\beq\label{cc}
= \frac{e}{2m_p}\,\langle n|(\hat{\p} + \frac{\bk}{2}) \exp(i\bk
\br/2)+ \exp(i\bk \br/2)\hat{\p}|0\rangle\, \\ =
\frac{e}{m_p}\,\langle n|\hat{\p} \exp(i\bk \br/2)|0\rangle\,;
\eeq
here $\hat{\p}$ acts on the relative coordinate $\br$.

At first let us take as intermediate states $|n\rangle$ in the
corresponding nuclear Compton amplitude just plane waves,
eigenstates of $\hat{\p}$. In this way we take into account all
the states with $l\neq 0$, which are free ones in our zero range
approximation, and in addition the $^3S_1$ wave function in the
free form $\psi_p(r)=\sin pr/pr$ (the deviation of the $^3S_1$
wave function from the free one will be considered below). Then,
with the zra deuteron wave function
\begin{equation}\label{gs}
\psi_0(r)\,=\,\sqrt{\frac{\varkappa}{2\pi}}\,\frac{\exp(-\varkappa
r)}{r}\,,
\end{equation}
the only matrix element entering the amplitude is
\begin{equation}\label{me}
\langle \psi_0\,|\exp(\pm
i\bk\br/2)|\p\rangle\,=\,\frac{\sqrt{8\pi\varkappa}}
{(\p\pm\bk/2)^2\,+\,\varkappa^2}\,.
\end{equation}
In this way the amplitude itself simplifies to
\begin{eqnarray}\label{ms2}
M_{mn}^{(2)}&=&\left(\frac{e}{2m_p}\right)^2
2\varkappa\omega\!\!\int\frac{d\p}{\pi^2}
\left\{\frac{\mu_p}{[(\p-\bk/2)^2+\varkappa^2]^2}
+\frac{\mu_n}{[(\p-\bk/2)^2+\varkappa^2][(\p+\bk/2)^2+\varkappa^2]}\right\}
\nonumber\\
&& \times\frac{\,2p_m\,i\,\epsilon_{nrs}k_r s_s\,
-\,2p_n\,i\,\epsilon_{mrs}k_r s_s}
{\omega^2\,-\,(p^2+k^2/4+\varkappa^2)^2/m_p^2}\,.
\end{eqnarray}

Due to the account for the motion of the system as a whole in the
intermediate states, this expression differs from the
corresponding one from our previous paper \cite{mkp} in two
respects. First, in \cite{mkp} the operator $\p_p$ in (\ref{cc})
was identified with $\p$. Thus, therein instead of $2p_{m,n}$ in
the analogue of the present formula (\ref{ms2}), we obtained
$(2p\,-\,k/2)_{m,n}$.  At present, in (\ref{ms2}) the term
proportional to $\mu_n$ is an odd function of $\p$, and therefore
vanishes after integration over $d\p$. Second, in the denominator
the energy difference has acquired the contribution $k^2/4m_p$\,,
which is the kinetic energy of the $pn$ system as a whole, and
thus $(p^2+\varkappa^2)/m_p$ has transformed into
$(p^2+k^2/4+\varkappa^2)/m_p$.

Now we substitute (\ref{ms2}) into (\ref{f}) and take the integral
over $\omega$ under the condition $\omega\gg\varkappa^2/m $. For
the relative correction to the hf structure we obtain
\begin{eqnarray}\label{relint2}
\Delta_{in}^{(2)}&=& \frac{2\alpha\varkappa\mu_p m_e}{\pi^4\mu_d
m_p}\!\!\int\!\!\int\frac{d\p\, d\bk}{k^4}\frac{\p\bk}
{[(\p-\bk/2)^2+\varkappa^2]^2}\left[\frac{m_p}
{p^2+k^2/4+\varkappa^2}-\frac{3}{2k}\right]\,.
\end{eqnarray}
The result of integration  over $\p$ and then over $\bk$ reads
\begin{equation}\label{di2}
\Delta_{in}^2=\,\alpha\,\frac{\mu_p}{\mu_d}\,\frac{m_e}{\varkappa}\,-
\,\frac{6\alpha}{\pi}\,\frac{\mu_p}{\mu_d}\,\frac{m_e}{m_p}\,\ln{\frac{m_p}{\varkappa}}\,
.
\end{equation}
The logarithmic contribution here originates from  integration  of
the term $3/2k$ in the square brackets in (\ref{relint2}) over the
range $\varkappa^2/m_p\ll k\ll \varkappa$. The result (\ref{di2})
differs from the corresponding one of \cite{mkp} by a term
proportional to $\mu_p+\mu_n$, which is relatively small
numerically. It is only natural that our present account for the
motion of the proton-neutron system as a whole in the intermediate
states results in the correction proportional to $\mu_p+\mu_n$.

Let us calculate now the correction $\Delta_{in}^{(3)}$
corresponding to the effect of deviation of the intermediate
$^3S_1$ wave function $\Psi_p(r)$ from the free one. In the zra
$\Psi_p(r)$ reads \beq \Psi_p(r) = \frac{\sin pr}{pr}
-\,\frac{1}{\varkappa + i
p}\,\frac{\exp(ipr)}{r}\,=\frac{\varkappa\sin pr -p\cos
pr}{pr(\varkappa+ip)}\, . \eeq It follows, for instance, from the
orthogonality to the deuteron wave function (\ref{gs}). We use
below the function \beq \rho_p(r_1,r_2)=\Psi_p(r_1)\Psi^*_p(r_2)-
\psi_p(r_1)\psi^*_p(r_2) = \frac{p\cos p (r_1+r_2)-\varkappa\sin
p(r_1+r_2)}{(\varkappa^2 +p^2 )pr_1r_2}\, . \eeq
 Then, after integration over $\omega$ the expression for $\Delta_{in}^{(3)}$
reads
\begin{eqnarray}\label{relint3}
&&\Delta_{in}^{(3)}= \frac{4\alpha(\mu_p+\mu_n) m_e}{\pi^3\mu_d
m_p}\!\!\int\!\!\int dk\,dp\,p^2 \!\!\int\!\!\int d\br_1d\br_2\,
\psi_0(r_1) \psi_0(r_2)\,\rho_p(r_1,r_2)\nonumber\\
&&\times\frac{\sin kr_1}{kr_1} \left[\frac{\sin
kr_2}{kr_2}-\frac{(1+\varkappa r_2)}{(kr_2)^2}\left(\frac{\sin
kr_2}{kr_2}-\cos kr_2\right)\right] \left[\frac{m_p}
{p^2+k^2/4+\varkappa^2}-\frac{3}{2k}\right].
\end{eqnarray}
The integral over $p$ is
\begin{eqnarray}
\int_0^\infty dp\,\frac{p^2\,\rho_p(r_1,r_2)
}{p^2+k^2/4+\varkappa^2}= \frac{2\pi}{r_1 r_2 k^2}\left[
(Q+\varkappa)\exp[-Q(r_1+r_2)]-2\varkappa\exp[-Q(r_1+r_2)]\right]
\,,
\end{eqnarray}
where $Q=\sqrt{\varkappa^2+k^2/4}$. Now we integrate
(\ref{relint3}) over $r_1$, $r_2$, and then over $k$. The final
result for the discussed correction reads:
\begin{equation}\label{di3}
\Delta_{in}^{(3)}=\,-\,\alpha\,
\frac{\mu_p+\mu_n}{\mu_d}\,\frac{m_e}{\varkappa}\,\frac{1}{3}\,(2-2\ln
2) +
\,\frac{3\alpha}{\pi}\,\frac{\mu_p+\mu_n}{\mu_d}\,\frac{m_e}{m_p}\,\ln{\frac{m_p}{\varkappa}}\,
.
\end{equation}
Again, it is only natural that due to common selection rules the
contribution of the $^3S_1$ intermediate state is proportional to
$\mu_p+\mu_n$. Let us note also that the first term in (\ref{di3})
is additionally suppressed by a small numerical factor $(2-2\ln
2)/3 = 0.20$.

At last, we wish to come back to the effect due to a finite
distribution of the deuteron charge and magnetic moment. The
Zemach correction $\Delta_{f}$ (\ref{zem}) can be also easily
derived in the present approach. Using the identity
$$\langle\psi_0\,|\hat{\p}\exp(
i\bk\br/2)|\psi_0\rangle\,=\frac{\bk}{4}\langle \psi_0\,|\exp(
i\bk\br/2)|\psi_0\rangle\, , $$ we obtain the corresponding
amplitude:
\begin{eqnarray}\label{zem1}
M_{mn}^{f}&=&\left(\frac{e}{2m_p}\right)^2 (\mu_p+\mu_n)\,\omega
\left[F^2(k)-1\right]\frac{\,k_m\,i\,\epsilon_{nrs}k_r s_s\,
-\,k_n\,i\,\epsilon_{mrs}k_r s_s}
{\omega^2\,-\,(p^2+k^2/4+\varkappa^2)^2/m_p^2}\,,
\end{eqnarray}
where $F(k)$ is the deuteron form factor in the zero-range
approximation
\begin{eqnarray}\label{ff}
F(k)=\langle \psi_0\,|\exp(i\bk\br/2)|\psi_0\rangle\,=\,
\frac{4\varkappa}{k}\arctan\frac{k}{4\varkappa}\,.
\end{eqnarray}
Let us note that in our approximation both electric and magnetic
form factors, which in the present case enter the convection
current and spin current matrix elements respectively, coincide
and equal $F(k)$.

The integration over $\omega$ leads to the following result for
the relative correction $\Delta_f$
\begin{eqnarray}\label{zem2}
\Delta_f=\frac{8\alpha(\mu_p+\mu_n)m_e}{\pi\mu_d }\int^\infty_0
\frac{dk}{k^2}\, [F^2(k)-1]=\, \,-\,\alpha\,
\frac{\mu_p+\mu_n}{\mu_d}\,\frac{m_e}{\varkappa}\,\frac{1}{3}\,(1+2\ln2).
\end{eqnarray}
There is no logarithmic term in $\Delta_f$ since $F^2(k)-1 \sim
k^2$ for $k \ll \varkappa$. In fact, the result (\ref{zem2})
agrees with (\ref{zem}) since to our accuracy $\mu_p+\mu_n =
\mu_d$.

The corrections (\ref{di2}), (\ref{di3}), (\ref{zem2}) combine
into a compact result
\begin{equation}\label{c}
\Delta_{c}=\,-\,\alpha\,\frac{\mu_n}{\mu_d}\,
\frac{m_e}{\varkappa}\,- \,\frac{3\alpha}{\pi}\,
\frac{\mu_p-\mu_n}{\mu_d}\,\frac{m_e}{m_p}\,\ln{\frac{m_p}{\varkappa}}\,.
\end{equation}

Let us notice that its logarithmic part coincides with the
corresponding logarithmic term in \cite{mkp} (see formula (27)
therein). This is quite natural: the log is dominated by small
$k$, so that extra power of $k$, which arises from the recoil of
the $pn$ system as a whole, cannot influence it.

The leading term in (\ref{c}) coincides with the result of Ref.
\cite{low}\footnote{We are sorry for misquoting the result of Ref.
\cite{low} in our paper \cite{mkp}.}. However, we could not find
any correspondence between our arguments and those of Ref.
\cite{low}. In particular, it is stated explicitly in Ref.
\cite{low} that the motion of the intermediate $pn$ system as a
whole is neglected therein.

\section{Discussion of results}

Our total result for the nuclear-structure corrections to the
deuterium hf structure, comprising all the contributions,
(\ref{de}), (\ref{di}), (\ref{di1}), (\ref{c}), is
\begin{equation}
\Delta=\,-\,\alpha\, \frac{\mu_n}{\mu_d}\,\frac{m_e}{\varkappa}\,
-\,\frac{3\alpha}{\pi}\,
\frac{\mu_p\,-\,\mu_n}{\mu_d}\,\frac{m_e}{m_p}\,\ln{\frac{m_p}{\varkappa}}
\end{equation}
\[
+\,\frac{3\alpha}{8\pi}\,\frac{(\,\mu_p\,-\,\mu_n\,)^2}{\mu_d}\,
\frac{m_e}{m_p}\,\ln{\frac{m_p}{\varkappa}}
\]
\[
 +\,\frac{3\alpha}{8\pi}\,\frac{1}{\mu_d}\,
(\,\mu_d^2\,-\,2\mu_d\,-\,3)\,\frac{m_e}{m_p}\,
\ln{\frac{\varkappa}{m_e}}\,
+\,\frac{3\alpha}{4\pi}\,\frac{1}{\mu_d}\,
(\,\mu_p^2-\,2\,\mu_p\,-\,3\,+\,\mu_n^2\,)\,\frac{m_e}{m_p}\,
\ln{\frac{m_{\rho}}{\varkappa}}\,.
\]
Numerically this correction to the hf splitting in deuterium
constitutes
\begin{equation}
\Delta \nu\,=\,50\, \mbox{kHz}.
\end{equation}

It should be compared with the lacking $45$ kHz (see (\ref{dif})).
We believe that the agreement is quite satisfactory if one recalls
the crude nuclear model (zra) used here; in particular the
deuteron form factors, as calculated in the zra, are certainly
harder than the real ones, and thus in zra the negative Zemach
correction is underestimated.

Let us mention here that in a recent paper \cite{mart} elastic
contributions and the Zemach effect were considered in a quite
different theoretical technique, but with some phenomenological
desription of the deuteron form factors. The result is smaller
than the corresponding part of ours by 13 kHz.

Clearly, the nuclear effects discussed are responsible for the
bulk of the difference between the pure QED calculations and the
experimental value of the deuterium hf splitting. The calculation
of this correction, including accurate treatment of nuclear
effects, would serve as one more sensitive check of detailed
models of deuteron structure.

\section*{Acknowledgements}
We are extremely grateful to J.L. Friar for insistently attracting
our attention to the role of the $^3S_1$ positive-energy state in
the inelastic contribution. The investigation was supported in
part by the Russian Foundation for Basic Research through Grants
Nos. 03-02-17612, 03-02-16510, and through Grant for Leading
Scientific Schools.

\end{document}